\documentstyle[12pt]{article}
\title{Measurability of Electromagnetic Field:\\
        Model and Path Integral Methods\thanks{Published in Phys. 
        Lett. A 188, 249-255, 1994.}}
\author{Horst v. Borzeszkowski\\ 
{\small Institut f\"ur Theoretische Physik}\\
{\small Technische Universit\"at Berlin, 
Hardenbergstr. 36, D-10623 Berlin, Germany}
\and Michael B. Mensky\\ 
{\small P.N.Lebedev Physical Institute, Leninsly prosp.43, 117924 
Moscow, Russia}} 
\date{}

\newcommand{\lo}{\raisebox{- .8 ex}
     {$\stackrel{\textstyle <}{\sim}$}}

\newcommand{\cl}{\rm class}

\newcommand{\al}{\alpha}

\newcommand{\be}{\begin{equation}}
\newcommand{\ee}{\end{equation}}
\newcommand{\ba}{\begin{eqnarray}}
\newcommand{\ea}{\end{eqnarray}}
\newcommand{\ban}{\begin{eqnarray*}}
\newcommand{\ean}{\end{eqnarray*}}

\newcommand{\om}{\omega}
\newcommand{\Om}{\Omega}
\newcommand{\Dx}{\Delta x}
\newcommand{\dEo}{\delta E_{\mbox{opt}}}
\newcommand{\Qo}{Q_{\mbox{opt}}}
\begin{document}
\maketitle
\begin{abstract}
The problem of the measurability of the electromagnetic field is 
investigated 1)~in the framework of the abstract 
restricted-path-integral method, and 2)~by explicitly accounting 
the action of the field onto the meter and its back reaction. The 
meaning of the previously obtained results as well as of the 
classical results of Bohr and Rosenfeld are made 
clear. The restricted-path-integral method with integration over 
field configurations is shown to give an estimation on the 
measurability of the field by any device not disturbing the 
measured field (in the process of measurement) more than by the 
measurement error. Such method of measurement is necessary 
for the control of the field in electronic devices. 
\end{abstract}
\section{Introduction}
The problem of measurability of the electromagnetic field was 
considered by Landau and Peierls (LP) in 1931~\cite{LPpaper}. 
Then the results of LP were revised by Bohr and Rosenfeld (BR) in 
1933~\cite{BRpaper}. In short, the difference between the two works is 
in that LP tried to prove the existence of  an absolute 
restriction on the measurability of the field, while BR argued that no 
absolute restriction exists. According to LP, no construction of the 
measuring device can provide a measurability beyond this absolute 
limit. BR argued that the measurability may be made arbitraryly 
small by choosing a sufficiently large charge of the test body. 

In the papers \cite{em-meas-AnPh,em-meas-TMF} and the book 
\cite{book3} the restricted-path-integral (RPI) method has been applied 
to the problem of the measurability of the electromagnetic field. The 
analysis in the framework of this method gave some estimation on 
the uncertainty $\delta E$ of the measurement output that does not 
always coincide with the previously known results. In the present 
paper we analyse this estimation in more detail applying a more 
explicit model of measurement. 

Our final conclusion is that the uncertainty $\delta E$ obtained 
by the RPI method gives the limits in which the measurement may 
be considered as undisturbing. Besides, more strong 
limitations for undiscturbing measurements are obtained in the 
case when the space and time dimensions of the measurement region 
satisfy the inequality $l<c\tau /137$ or $l>c\tau$. 

\section{Restricted-Path-Integral Method}\label{RPI}
The restricted-path-integral method \cite{book3} has been applied 
to the measurement of the electromagnetic field in 
\cite{em-meas-AnPh,em-meas-TMF} (see also \cite{book3}). We shall give 
here a short exposition of the method in its field-theoretic 
version appropriate for our aim. 

The  starting point is a functional integral on field 
configurations (which is often called also path integral). Let a 
field $\Phi$ be considered, and the action functional for this 
field be of the form
\be
S[\Phi]=\int_{\Om}d^{4}x\,L(\Phi,\partial\Phi)
\label{1e}\ee
where the integral is taken over a space-time region $\Om$. Then 
the dynamics of the quantum field is described by the amplitude 
equal to the integral over field configurations. For the dynamics in 
the space-time region $\Om$ the integral must be taken over 
field configurations in this region:          
\be 
U=\int d [\Phi]\,\exp(iS[\Phi]),
\label{2e}\ee

Measuring the field has influence on its dynamics that may be 
expressed by restricting the path integral. The restriction is 
determined by the information supplied by the measurement. Let 
the measurement give an output $\al$. Then the corresponding 
information may be expressed by some weight functional 
$w_{\al}[\Phi]$, small (or equal to zero) for all field 
configurations $[\Phi]$ incompatible with the information given 
by the measurement. In this case field configurations in the path 
integral should be weighted by the functional:   
\be 
U_{\al}=\int d [\Phi]\,w_{\al}[\Phi]\,\exp(iS[\Phi]).
\label{3e}\ee
The square modulus of the corresponding amplitude
$$
P_{\al}=|U_{\al}|^2
$$
gives a probability density for different measurement outputs $\al$. 

Applying this scheme of consideration to the measurement of the 
electromagnetic field strength one should take the action for 
this field and the path integral over its configurations 
\cite{em-meas-AnPh,em-meas-TMF,book3}. The unrestricted 
functional integral (\ref{2e}) has in this case the form 
\cite{Itz-Zub}:
\be 
U=\int d [A]\,\delta(\partial_{\mu}A^{\mu})
\exp\left[-\frac{i}{4}\int 
d^{4}x\,(\partial_{\mu}A_{\nu}-
\partial_{\nu}A_{\mu})(\partial^{\mu}A^{\nu}-\partial^{\nu}A^{\mu})\right].
 \label{4e}\ee
It is seen from this equation that the symbol $d [\Phi]$ in 
Eq.~(\ref{2e}) must be specified as $d 
[A]\,\delta(\partial_{\mu}A^{\mu})$ for the electromagnetic 
field. Expressing the field action through the field strength, 
one has
\be 
U=\int d [A]\,\delta(\partial_{\mu}A^{\mu}) 
\exp\left(-\frac{i}{2}\int d^{4}x\,({\bf H}_A^2-{\bf E}_A^2) \right). 
\label{5e}\ee

The preceding formulas are valid for the field in the absence of 
any measurement. Consider now the situation when the measurement 
of the field strength in the region $\Om$ is performed to give 
the result presented by the configurations $\vec{H}(x)$, 
$\vec{E}(x)$. This supplies an information about the field which 
can be characterized by the weight functional 
\be 
w_{[\vec{H},\vec{E}]}[A]=\exp\left[ 
-\frac{1}{\Om}\int d ^{4}x\,\left( 
\frac{({\bf H}_A-\vec{H})^2}{\Delta H^2} 
+\frac{({\bf E}_A-\vec{E})^2}{\Delta E^2} 
\right) 
\right]  
\label{6e}\ee 
where $\Om$ is the (four-dimensional) volume of the space-time 
region $\Om$ where the measurement is arranged. 

Indeed, this functional is almost equal to unity for the field 
configurations close to $\vec{H}(x)$, $\vec{E}(x)$ and it is almost 
zero otherwise. ``Being close'' is understood here in the sense of 
the square average. The functional decreases in $e$ times when 
the square average deviation of the magnetic field becomes larger 
than $\Delta H$ or/and the deviation of the electric field becomes 
larger than $\Delta E$. Therefore, the functional (\ref{6e}) 
describes the packet of configurations, which corresponds to the 
field measurement giving the output $[\vec{H},\vec{E}]$. 

The corresponding restricted (weighted) path integral has the 
form 
\ba 
\lefteqn{U_{[\vec{H},\vec{E}]}= 
\int d [A]\,\delta(\partial_{\mu}A^{\mu}) 
\exp\biggl[-\frac{i}{2}\int d^{4}x\,({\bf H}_A^2-{\bf E}_A^2)}\nonumber\\ 
&&-\frac{1}{\Om}\int d ^{4}x\, 
\left(\frac{({\bf H}_A-\vec{H})^2}{\Delta H^2} 
+\frac{({\bf E}_A-\vec{E})^2}{\Delta E^2} 
\right) 
\biggr].  
\label{7e}\ea
This integral gives an amplitude describing the measurement of 
electromagnetic field. The probability distribution over all 
possible measurement outputs (all possible field configurations 
$\vec{H}(x)$, $\vec{E}(x)$) is provided by the square modulus of 
the amplitude: 
$$ 
P_{[\vec{H},\vec{E}]}=|U_{[\vec{H},\vec{E}]}|^2
$$
The calculation shows \cite{em-meas-AnPh,em-meas-TMF,book3} that 
this probability distribution has the form
\be 
P_{[\vec{H},\vec{E}]}= 
\exp\left[-\frac{2}{\Om}\int_{\Om} d^{4}x\left( 
\frac{(\vec{H}-\vec{H}_{\cl})^2}{\Delta H^2+\frac{4}{\Om^2\Delta H^2}} 
+\frac{(\vec{E}-\vec{E}_{\cl})^2}{\Delta E^2+\frac{4}{\Om^2\Delta E^2}} 
\right)\right]. 
\label{15e}\ee

It follows from this distribution 
\cite{em-meas-AnPh,em-meas-TMF,book3} that the output of 
measurement $[\vec{H},\vec{E}]$ may differ from the classical 
configuration in such a way that the square average deviations 
\ba 
\|\vec{H}-\vec{H}_{\cl}\|^2&=& 
\frac{1}{\Om}\int_{\Om} d^{4}x\,(\vec{H}-\vec{H}_{\cl})^2,\nonumber\\ 
\|\vec{E}-\vec{E}_{\cl}\|^2&=& 
\frac{1}{\Om}\int_{\Om} d^{4}x\,(\vec{E}-\vec{E}_{\cl})^2 
\label{16e}\ea
be not too large, namely satisfy the following inequalities:
\be 
\|\vec{H}-\vec{H}_{\cl}\|^2\,\lo\,\delta H^2, 
\quad 
\|\vec{E}-\vec{E}_{\cl}\|^2\,\lo\,\delta E^2, 
\label{17e}\ee
with
\be 
\delta H^2 =\Delta H^2+\frac{4}{\Om^2\Delta H^2}, 
\quad 
\delta E^2 =\Delta E^2+\frac{4}{\Om^2\Delta E^2}. 
\label{18e}\ee

Two qualitatively different regimes of the measurement may be 
considered. If 
$$ 
\Delta H\gg\sqrt{\frac{2}{\Om}},
\quad \Delta E\gg
\sqrt{\frac{2}{\Om}}, 
$$ 
then the second term in the right-hand side of each of the 
formulas (\ref{18e}) is negligible, so that (\ref{18e}) takes 
form 
\be 
\delta H=\Delta H,\quad \delta E=\Delta E. 
\label{19e}\ee
If, on the contrary, 
$$ 
\Delta H\ll\sqrt{\frac{2}{\Om}},
\quad \Delta E\ll
\sqrt{\frac{2}{\Om}}, 
$$ 
then the first terms in (\ref{18e}) become small, so that 
\be 
\delta H =\frac{2}{\Om\Delta H},\quad 
\delta E =\frac{2}{\Om\Delta E}. 
\label{20e}\ee
The first regime is a classical one because it completely 
corresponds to the classical theory of measurement. The second 
regime is essentially quantum. 

Returning to the general formula (\ref{18e}) one sees that the 
variance of the measurement results has the minimal value: 
\be 
\delta H_{\rm min}=
\delta E_{\rm min}=
\frac{2}{\sqrt{\Om}}. 
\label{21e}\ee
This corresponds to an optimal regime of measurement lying on 
the border between the classical and quantum ones. The existence of 
the minimum means that there is an absolute restriction on the 
measurability of the field. The limiting measurability is 
determined by the variance
\be 
\delta H_{\rm min}=\delta E_{\rm min}=\frac{2}{\sqrt{\tau l^3}}. 
\label{22e}\ee

In all preceding consideration natural units were used in which 
$\hbar=c=1$. In ordinary units one has for the absolute limit
\be 
\delta H_{\rm min}=\delta E_{\rm min}=2\sqrt{\frac{\hbar}{\tau l^3}}. 
\label{23e}\ee
 
\section{Explicit Account of Back Reaction}
The restricted-path-integral method used in Sect.~\ref{RPI} gives 
an absolute limit for the measurability of an electromagnetic field 
for a certain definition of the measurement (we shall discuss 
this definition in detail later, in Sect.~\ref{undisturb}). This 
means that the quantum measurement noise does not allow one to 
have information more precise than is expressed by 
Eq.~(\ref{23e}) about the value of the field strength. It is not 
clear from the phenomenological restricted-path-integral method 
what the origin of the quantum noise is and what factors do 
contribute to it. The rest of the paper is devoted to an analysis 
of these questions. 

We shall show that the measurement noise consists of two 
characteristic parts, the mechanical uncertainty of the probe 
body and a proper field of this body. It is the first part that 
has been taken into account in the well-known paper of Bohr and 
Rosenfeld \cite{BRpaper}. In the book \cite{Borz-Tred} this part has 
been found with the help of the restricted-path-integral method 
(but with integration over trajectories of a mechanical test 
body, not over field configurations). We shall remind the results 
of these works in the present section. 

Bohr and Rosenfeld derived their estimation for the measurability of 
an electric field, 
\be\label{BR}
\delta E_{\mbox{BR}} = \frac{\hbar c}{\Delta x\, c\tau \,Q}, 
\ee
by taking Heisenberg's uncertainty principle into account. For the 
measurement of the field the momentum of a probe charge must be 
found from observing its movement. However a precise localization 
of the charge in the observation prevents a precise determination 
of its momentum due to the uncertainty principle. The minimal 
possible error (\ref{BR}) in the estimation of the field results 
from optimization of the process. 

The same conclusion can be drawn \cite{Borz-Tred} if the 
observation of the measuring charge is considered with the help 
of the restricted-path-integral method. 

Let, for example, this charge be an oscillator with the 
frequency\footnote{Free charge can be considered as a special 
case $\om=0$.} $\om$ and the characteristic frequency of the 
measured motion is $\Om$. Then an optimal value of the 
measurement error may be shown \cite{book3} to be 
\be\label{MB1}
\Delta x = \left( \frac{\hbar}{m\tau |\Omega^2 - \omega^2|}\right)^{\frac12}
\ee
and the precision with which the force acting on the oscillator 
can be estimated is 
\be\label{MB2}
\delta F = \left( \frac{m\hbar|\Omega^2 - \omega^2|}{\tau 
}\right)^{\frac12}. 
\ee

Taking into account that $F=QE$ and multiplying the preceding 
formulas one has the same estimation for the uncertainty of the 
measured value of the field as in the paper of Bohr and Rosenfeld:
\be\label{MB}
\delta E_{\mbox{mech}} = \delta E_{\mbox{BR}} = 
\frac{\hbar c}{\Delta x\, c\tau \,Q}
\ee

This formula has a very simple and characteristic 
structure. Indeed, if one takes into account that $Q\Delta E$ is 
an uncertainty in the force and $\Delta p = Q\Delta E \,\tau $ is an 
uncertainty in the momentum acting on the measuring body then 
Eq.~(\ref{MB}) reduces to the uncertainty relation for this 
body,\footnote{We use an equality instead of an 
inequality meaning a limiting (optimal in quantum sense) regime of 
measurement.} $\Delta p \Delta x = \hbar$. 

Notice that the uncertainty discussed in this section is a 
consequence of the quantum uncertainties of the mechanical meter 
used for the measurement of the field. If one takes into account 
that this mechanical device must have a charge to interact with 
the field, the question arises about proper fields of the 
measuring body. 

The question about proper fields was discussed in literature and 
particularly in \cite{BRpaper}. It was concluded that proper 
fields do not prevent measurement because they may be calculated 
(i.e. they are systematic errors that can be taken into 
account). We shall however discuss this question in a new light. 
Namely, we shall consider the situation when these systematic 
errors are an obstacle for the aim of the measurement. This is, 
for example, the case when the measurement aims at controlling the 
field so that the observer would like to know what is the real 
value of the field, where the influence of the measurement is 
also taken into account. We shall see that it is just this setup 
of the problem that is characteristic for the 
restricted-path-integral approach. 

\section{Accuracy with which the Measurement is Undisturbing}\label{undisturb}
In the paper of Bohr and Rosenfeld \cite{BRpaper} the uncertainty 
(\ref{MB}) is considered as the only obstacle for measurability, 
and from some point of view it is. However, one could introduce 
a further characteristic for the measurement, $\delta E$, including 
an additional field arising during the measurement, into this 
entity. 

The motivation for this is evident. In some situations the aim 
of the measurement is not the estimation of what the filed could be 
if the measurement is not actually performed, 
but the estimation of what the field really was, with all actual 
circumstances accounting, among them the measurement itself. If 
one knows the entity $\delta E$ one can be sure that the field 
was actually equal to the measurement outcome, E, with the 
precision $\Delta E$. 

We have for this new characteristic the evident formula
\be\label{variety}
\delta E = \delta E_{\mbox{mech}} + E_{\mbox{meas}}
\ee
where $E_{\mbox{meas}}$ is the complete field created by 
measuring bodies. 

Let our measuring body has a charge $Q$ and the measurement is 
arranged in the region of the size $l$. Accept the 
Coulomb formula 
\be\label{coul}
E_{\mbox{meas}} = \frac{Q}{l^2}
\ee
as the simplest estimation for the field of such 
a body in a typical point of the measurement region. Then 
\be\label{variety-coul}
\delta E = 
\frac{\hbar c}{\Delta x\, c\tau \,Q}
+ \frac{Q}{
l^2}.
\ee

We see from Eq.~(\ref{variety-coul}) that the complete (with 
the additional fields) measurement uncertainty $\delta E$ 
depends (for given dimensions of the measurement region, $l$ and 
$\tau$) on the value of the charge $Q$ of the measuring body and 
the measurement error of the mechanical meter, $\Delta x$. We 
should choose these parameter in such a way that the uncertainty 
$\delta E$ be minimal. This will give an absolute limit 
on the measurability of the field in an undisturbing regime. 

Let us consider first the choice of the charge $Q$ for a fixed error 
$\Delta x$. It is evident that the expression 
(\ref{variety-coul}) has a minimum achieved for the charge equal 
to 
\be\label{opt-charge} 
Q_{\mbox{opt}}^2 =\hbar c\,\frac{l^2}{\Delta x\, c\tau }. 
\ee
For this value of the charge we have the uncertainty equal to 
\be\label{uncert-opt}
\delta E_{\mbox{opt}} = 
2\sqrt{\frac{\hbar c}{\Delta x\, c\tau \,l^2}}.
\ee

It is seen now that one should increase $\Delta x$ to diminish the 
uncertainty $\dEo$. However the error $\Delta x$ in measurement 
of the position of the measuring body cannot be more than the 
size $l$ of the measuring region: 
\be\label{dxl}
\Delta x \leq l. 
\ee
Taking\footnote{We shall consider in the next sections the 
situations when this is impossible.} $\Dx=l$ we have for the 
minimum possible uncertainty the following estimation:
\be\label{uncert-abs}
\delta E_{\mbox{abs}} = 
2\sqrt{\frac{\hbar c}{c\tau \,l^3}}.
\ee
It is evident that this estimation concides with the estimation 
(\ref{23e}) found by the restricted-path-integral method. 

The formula (\ref{uncert-abs}) and the analysis leading to this 
formula makes more clear the sense of the estimation (\ref{23e}) 
and, more generally, of the estimations found by the 
restricted-path-integral method. We see that this method gives 
a restriction for undisturbing measurement. If the output of the 
measurement is $E$ and the uncertainty of the output found by the 
restricted-path.integral method is $\delta E$, then we know that 
the field really was in the limits of the interval $[E-\delta E, 
E+\delta E]$, even with the fields of measuring bodies taken into 
account. 

\section{Accounting quantization of charge}\label{qu-charge}
Let us consider now more attentively the case when the choice $Q=\Qo$ and $\Dx=l$ 
is possible so that one is led to the above estimation. The 
problem is that, 
due to the relation (\ref{opt-charge}), enlarging of $\Dx$ leads to 
diminishing of $\Qo$. Choosing $\Dx=l$ we determine some value 
for the charge $\Qo$ and we should be sure that this charge is 
feasible, i.e., is larger than the charge of an electron, 
$$
\Qo > e.
$$
Otherwise we should take the value for $\Dx$ less than $l$. 

The value of $l/\Dx$ leading to the value $e$ for $\Qo$ is 
$$
\frac{l}{\Dx}=\frac{1}{137}\frac{c\tau}{l}
$$
where $1/137$ is the (approximate) value of the fine structure 
constant, $e^2 / \hbar c$. Therefore we should in fact choose for 
$l/\Dx$ not the value 1 as was done above, but 
$$
\frac{l}{\Dx}=\max\left( 1, \frac{1}{137}\frac{c\tau}{l} \right).
$$

We see therefore that the above consideration leading to 
Eq.~(\ref{uncert-abs}) is valid only for $l/c\tau$ greater 
than $1/137$. If the opposite inequality
$$
\frac{l}{c\tau}\leq \frac{1}{137}. 
$$
is valid, we are led to the estimation
\be\label{uncert-abs-e}
\delta E_{\mbox{abs}} = 
\frac{2e}{l^2}=
\frac{1}{6}\frac{\sqrt{\hbar c}}{l^2}
\ee
where $1/6$ is accepted as an approximate value for 
$2/\sqrt{137}$. This estimation is valid for the case when 
an optimal value of the measuring charge is $e$. 

In the latter formula the discrete structure of matter is taken into 
account which is impossible to do in the framework of the 
phenomenological restricted-path-integral approach.

\section{Accounting causal relations}

Let us consider a measurement arranged in the space region of 
the size $l$ during the time $\tau$ but with the inequality 
\be\label{acausal-region}
l > c\tau
\ee
valid for these parameters. In this case the measurement region 
is ``acausal'' in the following sense. In the space region of the 
size $l$ some smaller subregions (of the size say $\lambda$) 
exist that, during the time of measurement, $\tau$, cannot be 
connected by a sublight signal. Any two events which occured in two 
such subregions cannot have causal influence on each other. Such 
subregions are causally disconnected. Therefore, the whole 
space-time region where the measurement is arranged is in this 
sense ``acausal''. 

Bohr and Rosenfeld considered such acausal measurement regions. 
This is why we shall also consider them. We shall see that the 
estimations (\ref{uncert-abs}), (\ref{23e}) are incorrect in this 
case too.\footnote{Notice however that we consider another 
characteristic of the measurement here than in the paper of BR so 
that the direct comparison of our results with theirs is 
impossible.}

The formula (\ref{MB}) is valid also in the case of acausal 
region. However Eqs.~(\ref{uncert-abs}), (\ref{23e}) cannot be 
used for such a region. Instead, $l=c\tau$ should be 
substituted in these formulas. The reason is the following. 

The entity $\Omega$ in Eq.~(\ref{6e}) and the subsequent formulas is 
the four-volume of a region where an ``integral'' measurement 
is performed that cannot be reduced to measurements arranged in 
smaller regions. If an ``elementary'' measurememnt is arranged in the 
region of the volume $\omega =\lambda^3c\tau$ then just this 
volume should be substituted in a denominator of the exponent 
(\ref{6e}) (but the integration should be performed over the whole 
region $\Omega$). The resulting estimation for the variety of the 
measurment outputs is 
\ba
\Delta E_{\rm opt}^2= 
\Delta H_{\rm opt}^2&=&\frac{2}{\om} \\
\delta E_{\rm min}=\delta H_{\rm min}
&=&2\sqrt{\frac{\hbar}{\tau \lambda^3}}. 
\ea

In the case $l<c\tau$ an ``integral'' measurement is possible. If 
however $l>c\tau$, the measurement procedure decomposes into a 
series of independent procedures arranged in causal parts of the 
whole region. The maximal size of such a part is $l=c\tau$. This 
is why we should substitude $\om =\lambda(c\tau)^3 =(c\tau)^4$ in 
this case. This leads to 
\be\label{acausal-uncert}
\delta E_{\rm min}
=2{\frac{\sqrt{\hbar c}}{(c\tau)^2}}. 
\ee

The argument concerning the incorporation of the field of 
measuring bodies should also be changed because the field in each 
of the causal components of an acausal region must be considered 
separately. Let us denote the charge of the measuring bodies in 
such a subregion by $q$. Then instead of Eq.~(\ref{variety-coul}) 
we have 
\be\label{variety-coul-elem}
\delta E = 
\frac{\hbar c}{\Delta x\, c\tau \,q}
+ \frac{q}{\lambda^2}=
\frac{\hbar c}{\Delta x\, c\tau \,q}
+ \frac{q}{(c\tau)^2}.
\ee
The optimization of this formula (with accounting of $\Delta 
x<\lambda =c\tau$) gives 
\be
\delta E_{\rm abs}
=2{\frac{\sqrt{\hbar c}}{(c\tau)^2}}. 
\ee
Thus both methods give the same estimation of $\delta E$ in an 
acausal case too, but this estimation differs from that obtained 
for a causal region. 

\section{Discussion}
In the present work we considered the measurement of the 
electric field in the framework of two different approaches 
and compared the corresponding conclusions. The first approach 
based on restricting path integrals is phenomenological. The 
second approach includes an explicit consideration of mechanical 
properties of a measuring body as well as its electrical charge. 
When applying the second approach, we took into consideration 
both 1)~the uncertainty relation for the measuring body as a 
mechanical system and 2)~its proper electric field as an obstacle 
for undisturbing measurement of an external field. 

The comparison of the results of both approachs allows us to conclude 
the following: 
\begin{itemize}
   \item The measurement uncertainty $\delta E$ obtained from the 
restricted-path-integral (RPI) approach determines the precision 
with which the measurement may be considered undisturbing. 
   \item The value (\ref{23e}), (\ref{uncert-abs}) for $\delta E$ 
which is obtained by the RPI method is an absolute restriction for the 
undisturbing measurement arranged in the region with the space 
dimension of the order of $l$ during time of the order of $\tau$. 
   \item The RPI method does not take into account the discrete 
structure of matter, in particular the fact that any charge is a 
sum of charges equal to the electron charge $e$. Taking into 
accoung this curcumstance one obtains in the case $l<c\tau 
/137$ the more strong restriction (\ref{uncert-abs-e}) for a 
undisturbing measurement (than by the RPI method). 
   \item In the case $l>c\tau$, i.e., in an acausal measurement 
region, the RPI method should be applied to each causal subregion 
separately. This leads to a more strong limitation 
(\ref{acausal-uncert}) than in a generic case. 
\end{itemize}

\begin{center}ACKNOWLEDGEMENT\end{center}

The work of one of the authors (MBM) on this paper was partly 
supported by Deutsche Forschungsgemeinschaft.

\end{document}